\documentclass[aps,prd,onecolumn,groupedaddress,showpacs,nofootinbib,amssymb
]{revtex4}
\usepackage[dvips]{graphicx}
\usepackage{amssymb}
\usepackage{amsmath}
\usepackage{graphicx}
\usepackage{amsfonts}
\usepackage{bm}

\begin{document}

\title{The Classical and Loop Quantum Cosmology Phase Space of Interacting Dark Energy and Superfluid Dark Matter}
\author{V.K. Oikonomou,$^{1,2,3}$\,\thanks{v.k.oikonomou1979@gmail.com}}
\affiliation{$^{1)}$ Department of Physics, Aristotle University of Thessaloniki, Thessaloniki 54124, Greece\\
$^{2)}$ Laboratory for Theoretical Cosmology, Tomsk State
University of Control Systems
and Radioelectronics (TUSUR), 634050 Tomsk, Russia\\
$^{3)}$ Tomsk State Pedagogical University, 634061 Tomsk, Russia }
\tolerance=5000

\begin{abstract}
In this paper we study in detail the phase space of a cosmological
system consisting of two coupled fluids, namely a dark energy
fluid coupled with a superfluid dark matter fluid. The dark matter
fluid is assumed to have a superfluid equation of state, hence it
is not pressureless and our aim is to find the impact of this
non-trivial equation of state on the phase space of the coupled
system. We shall use two theoretical contexts, namely that of
classical cosmology and that of loop quantum cosmology. In the
classical case, we investigated the existence and stability of
fixed points, and as we will show, no de Sitter fixed points
occur, however matter and radiation domination fixed points occur,
which are hyperbolic and unstable. We also show that there exist
limited sets of initial conditions for which singular solutions
occur in the phase space. With regard to the loop quantum
cosmology case, we demonstrate that stable de Sitter fixed points
exist, for some values of the free parameters of the theory, and
interestingly enough, for the same values, singular solutions
corresponding to general sets of initial conditions occur. To our
knowledge this feature does not occur so frequently in loop
quantum cosmological frameworks. We also demonstrate that
non-singular solutions corresponding to a general set of initial
conditions occur, however these occur when the dark matter
superfluid has negative pressure, so it is a rather physically
unappealing situation.
\end{abstract}


\maketitle

\section{Introduction}

The evolution of the Universe at present time seems to be driven
by dark forces of unknown nature and form, which currently are
known as dark energy and dark matter. The dark energy controls the
current acceleration of the Universe which was discovered in the
late 90's \cite{Riess:1998cb}, to a percentage of nearly $72\%$,
and it characterizes a negative pressure fluid, while dark matter
corresponds to the $24\%$ of the total energy density of the
Universe. With regard to dark matter, its presence is compelling
since the successful $\Lambda$-Cold Dark Matter ($\Lambda$CDM)
model relies on the existence of this dark component. Indeed, the
formation of galaxies and the galactic rotation curves assume the
presence of a non-interacting pressureless fluid component of our
Universe. Its nature is up to date still unknown, however there
exist several particle physics models which indicate that dark
matter is actually a non-interacting particle, see for example
\cite{Oikonomou:2006mh}, nevertheless no indication of the
particle nature of dark matter exists for the moment. One
appealing description for dark energy is provided by modified
gravity models
\cite{reviews1,reviews2,reviews3,reviews4,reviews5,reviews6}, with
the most popular description coming from $f(R)$ gravity. In fact,
it is possible to harbor both the early-time and late-time
acceleration eras of our Universe in the theoretical framework of
$f(R)$ gravity, see for example the model developed in
\cite{Nojiri:2003ft}. Since the dark sector still remains a great
mystery for theoretical cosmologists, there exists a research
stream in modern theoretical cosmology which assumes that both
dark energy and dark matter are modelled by some interacting
fluids, with the dark energy having a generalized equation of
state (EoS), see for example Refs.
\cite{Gondolo:2002fh,Farrar:2003uw,Cai:2004dk,Bamba:2012cp,Guo:2004xx,Wang:2006qw,Bertolami:2007zm,He:2008tn,Valiviita:2008iv,Jackson:2009mz,Jamil:2009eb,He:2010im,Bolotin:2013jpa,Costa:2013sva,Boehmer:2008av,Li:2010ju,Yang:2017zjs}.
The fluid cosmological description is frequently adopted in the
literature to explain several evolutionary features of the
Universe, see for example Refs.
\cite{Barrow:1994nx,Tsagas:1998jm,HipolitoRicaldi:2009je,Gorini:2005nw,Kremer:2003vs,Brevik:2018azs,Carturan:2002si,Buchert:2001sa,Hwang:2001fb,Cruz:2011zza,Oikonomou:2017mlk,Brevik:2017juz,Brevik:2017msy,Nojiri:2005sr,Capozziello:2006dj,Nojiri:2006zh,Elizalde:2009gx,Elizalde:2017dmu,Brevik:2016kuy,Balakin:2012ee,Zimdahl:1998rx,Odintsov:2018obx},
for an important stream of papers and reviews. A non-trivial
interaction between the dark sector fluids is motivated due to the
fact that the dark energy component of our Universe utterly
dominates after the galaxy formation period, during the matter
domination era, and this dominance continues until present time.
Another motivation to use a non-trivial interaction between the
dark fluids is the interdependence of dark energy and dark matter,
since evidence is provided that the energy density of dark matter
cannot be calculated without knowing the dark energy density
$\Omega_{DE}$, see Ref. \cite{Kunz:2007rk} where the dark energy
models degeneracy is discussed. Notably, interacting dark
energy-dark matter models are known to cause instabilities during
the primordial acceleration of the Universe, so this should be
kept in mind for an accurate model building. It is also useful to
note that in all interacting dark energy-dark matter models, the
luminous mass, that is the baryonic matter, must be uncoupled from
the dark sector fluids, because this would generate a fifth force
in the Universe, which is a rather unphysical effect, at least for
the time being.

In all the above mentioned interacting multifluid studies, the
dark matter fluid was assumed to be a pressureless perfect fluid,
however an interesting proposal appeared by the authors of
\cite{Berezhiani:2015bqa}, see also
\cite{Berezhiani:2015pia,Hodson:2016rck,Berezhiani:2017tth},
indicating that dark matter might be some sort of superfluid at
the galactic scale, and also at cosmological scales may be
described as an ordinary pressureless fluid. This interesting
theoretical idea is supported by the fact that the $\Lambda$CDM
model has several shortcomings at the galactic level, see for
example the review \cite{Tulin:2017ara}. Indeed, as was also
indicated in \cite{Berezhiani:2015bqa}, three major obstacles
challenge the $\Lambda$CDM model at galactic scales, namely, the
regularity of the galaxies and the corresponding Baryonic Tully
Fisher Relation, the dwarf galaxies of the Local Group, and the
tidal dwarfs recycled galaxies which emanate from the tidal
material created by merging spirals. As was shown in
\cite{Berezhiani:2015bqa}, a superfluid dark matter EoS at a
galactic level may actually provide useful insights towards the
understanding  galactic scale dynamics. To this end, in this paper
we shall investigate the dynamical evolution of a cosmological
system consisting of two interacting fluids, namely dark energy
with superfluid dark matter. We shall focus on the phase space
structure of the model, in order to investigate which are the
fixed points of such a system and examine their stability. Our
main interest is to see what effects would have a local superfluid
dark matter component on the dark energy fluid, and in effect on
the evolution of the Universe, in terms of fixed points and their
stability. Although dark matter at large scales seems to be better
modelled by a pressureless fluid, our interest is to check the
effect of a superfluid EoS at the phase space of interacting dark
energy dark matter fluids. Our results will provide hints about
the global effect of superfluid dark matter on the phase space of
the cosmological system. As for the dark matter fluid, we shall
assume that it will have a generalized EoS of polynomial form,
used frequently in research articles in the field
\cite{Odintsov:2018uaw,Odintsov:2018awm}. The structure of the
cosmological system consisting of dark energy and superfluid dark
matter motivates us to use specific appropriately chosen variables
in order to construct an autonomous dynamical system, which can
provide concrete information about the phase space structure. The
approach of using dynamical systems in cosmology is frequently
adopted in the literature
\cite{Odintsov:2018uaw,Odintsov:2018awm,Odintsov:2019ofr,Boehmer:2014vea,Bohmer:2010re,Goheer:2007wu,Leon:2014yua,Guo:2013swa,Leon:2010pu,deSouza:2007zpn,Giacomini:2017yuk,Kofinas:2014aka,Leon:2012mt,Gonzalez:2006cj,Alho:2016gzi,Biswas:2015cva,Muller:2014qja,Mirza:2014nfa,Rippl:1995bg,Ivanov:2011vy,Khurshudyan:2016qox,Boko:2016mwr,Odintsov:2017icc,Granda:2017dlx,Landim:2016gpz,Landim:2015uda,Landim:2016dxh,Bari:2018edl,Chakraborty:2018bxh,Ganiou:2018dta,Shah:2018qkh,Oikonomou:2017ppp,Odintsov:2017tbc,Dutta:2017fjw,Odintsov:2015wwp,Kleidis:2018cdx},
and it provides concrete results on the phase space structure of
several cosmological systems. We will examine the fixed points of
the cosmological system at hand and their stability and also we
investigate the dynamical evolution of the system. The results
indicate the existence of instabilities and of singular solutions.
Thus by using dynamical systems techniques, we demonstrate that
there exist singular trajectories in the phase space of the
cosmological system, which however correspond to a limited set of
initial conditions. This result is intriguing, so a question
naturally spring to mind, does these singular solutions exist even
if Loop Quantum Cosmology (LQC) effects are taken into account?
The LQC theoretical framework
\cite{LQC1,LQC3,LQC4,LQC5,Salo:2016dsr,Xiong:2007cn,Amoros:2014tha,Cai:2014zga,deHaro:2014kxa,Kleidis:2018plu,Kleidis:2017ftt},
is known to remove any classical singularities of cosmological
models, see for example \cite{Sami:2006wj} for a characteristic
example of this sort of behavior. Thus, we shall extend the
analysis performed in the classical case, and study the dynamical
system of the interacting dark energy and superfluid dark matter
fluids. As we demonstrate, in this case too, it is also possible
to construct an autonomous dynamical system, so the investigation
by using the dominant balance analysis becomes reliable. By using
well known techniques of autonomous dynamical systems, we find the
fixed points of the system, discuss their stability and we also
investigate whether singular solutions corresponding to global
initial conditions exist. To our great surprise, there exist
general singular solutions for some parameter values, and this is
our first time that we find such singular solutions in the context
of LQC. Intriguingly enough, for the same values of the free
parameters, there also exist stable de Sitter fixed points. Also
as we demonstrate, there exist general non-singular solutions too,
for different values of the free parameters, and we also question
the physical significance of the singular solutions, due to the
fact that these originate from a negative pressure dark matter
fluid. As a final task, we discuss the type of global finite-time
singularities that may occur even in the LQC system, by using the
classification performed firstly in Ref. \cite{Nojiri:2005sx}.

This paper is organized as follows: In section II we study the
classical coupled dark energy superfluid dark matter system, and
we investigate how to construct a polynomial autonomous dynamical
system. We also investigate the existence and stability of
hyperbolic fixed points and by using well-known techniques, we
investigate whether singular solutions exist. In section III we
perform the same analysis for the LQC coupled dark energy
superfluid dark matter system, and we emphasize on the existence
and stability of de Sitter fixed points. Also we investigate the
conditions under which singular solutions exist in the phase
space. Finally the concluding remarks appear in the end of the
paper.

Before starting, we need to discuss the geometric framework we
shall use in this work, which is a flat Friedmann-Robertson-Walker
(FRW) spacetime, with metric,
\begin{equation}\label{frw}
ds^2 = - dt^2 + a(t)^2 \sum_{i=1,2,3} \left(dx^i\right)^2\, ,
\end{equation}
where $a(t)$ is as usual the scale factor of the Universe. The
corresponding Ricci scalar is,
\begin{equation}\label{ricciscalaranalytic}
R=6\left (\dot{H}+2H^2 \right )\, ,
\end{equation}
with $H=\frac{\dot{a}}{a}$ denoting the Hubble rate of the
Universe.

\section{The Classical Cosmology Framework and Interacting Dark Energy with Superfluid Dark Matter}

We start off our analysis with the classical cosmological system
of dark energy-dark matter interacting fluids, with the dark
energy component having energy density denoted as $\rho_d$ and the
dark matter component having energy density $\rho_m$. The
cosmological equations in the flat FRW background (\ref{frw})
read,
\begin{equation}\label{flateinstein}
H^2=\frac{\kappa^2}{3}\rho_{tot}\, ,
\end{equation}
with $\kappa^2=8\pi G$, and $G$ being Newton's gravitational
constant, and also with $\rho_{tot}$ we denote the total energy
density of the cosmological system, which is equal to,
\begin{equation}\label{totealeeos}
\rho_{tot}=\rho_m+\rho_d\, .
\end{equation}
In the following we shall use a physical units system in which
$\hbar=c=1$. Due to the conservation of energy-momentum for the
interacting dark energy-dark matter system, we have the following
conservation of energy equations,
\begin{align}\label{continutiyequations}
& \dot{\rho}_m+3H(\rho_m+p_m)=Q\,
\\ \notag & \dot{\rho}_d+3H(\rho_d+p_d)=-Q\, ,
\end{align}
where with $p_m$ and $p_d$ we denote the pressure of the dark
matter and dark energy component respectively, hence we assume
that the dark matter component has non-zero pressure, the exact
form of which we present shortly. Also the parameter $Q$ in Eq.
(\ref{continutiyequations}) denotes the non-trivial interaction
between the dark sector fluids, the sign of which indicates which
fluid loses energy. Obviously, if $Q<0$ the dark matter sector
will lose energy and the dark energy sector gains, and the
converse occurs if $Q>0$. A phenomenologically motivated form of
the interaction term $Q$ is the following
\cite{CalderaCabral:2008bx,Pavon:2005yx,Quartin:2008px,Sadjadi:2006qp,Zimdahl:2005bk},
\begin{equation}\label{qtermform}
Q=3H(c_1\rho_m+c_2\rho_d)\, ,
\end{equation}
with $c_1$, $c_2$ being real constants of the same sign. By
differentiating Eq. (\ref{flateinstein}), and making use of the
continuity equations (\ref{continutiyequations}) we obtain,
\begin{equation}\label{derivativeofh}
\dot{H}=-\frac{\kappa^2}{3}\left( \rho_m+\rho_d+p_{tot}\right)\, ,
\end{equation}
with $p_{tot}$ being the total pressure, which is equal to
$p_{tot}=p_d+p_m$, since the dark matter component has non-zero
pressure too. Let us now assume that the superfluid dark matter
component has an EoS of the form $p_m\sim
\rho_m^3$\cite{Berezhiani:2015bqa}, and for the purpose of this
paper we assume it has the following form,
\begin{equation}\label{darkmattereos}
p_m=B\kappa^8\rho_m^3\, ,
\end{equation}
where $B$ is a dimensionless variable. Also, we assume that the
dark energy fluid obeys the generalized EoS, \cite{Nojiri:2005sr},
\begin{equation}\label{darkenergyeos}
p_d=-\rho_d-A\kappa^4\rho_d^2\, ,
\end{equation}
with $A$ being a real dimensionless constant. Having Eqs.
(\ref{flateinstein}), (\ref{continutiyequations}),
(\ref{derivativeofh}), (\ref{darkmattereos}) and
(\ref{darkenergyeos}) at hand, we can construct an autonomous
dynamical system by choosing the variables of the dynamical system
as follows,
\begin{equation}\label{variablesofdynamicalsystem}
x_1=\frac{\kappa^2\rho_d}{3H^2},\,\,\,x_2=\frac{\kappa^2\rho_m}{3H^2},\,\,\,z=\kappa^2H^2\,
,
\end{equation}
and we need to stress that all the variables of the dynamical
system are dimensionless. By using the variables $x_1$, $x_2$ and
$z$, the Friedmann equation (\ref{flateinstein}) becomes,
\begin{equation}\label{friedmannconstraint}
x_1+x_2=1\, ,
\end{equation}
to which we shall refer as ``Friedmann constraint'' hereafter. In
terms of the variables (\ref{variablesofdynamicalsystem}), the
interaction term (\ref{qtermform}) can be written as follows,
\begin{equation}\label{additionalterms}
\frac{\kappa^2Q}{3H^3}=3c_1x_2+3c_2x_1\, .
\end{equation}
By combining Eqs. (\ref{flateinstein}),
(\ref{continutiyequations}), (\ref{derivativeofh}),
(\ref{darkmattereos}), (\ref{darkenergyeos}),
(\ref{variablesofdynamicalsystem}) and (\ref{additionalterms}),
and also by using the $e$-foldings number $N$ as the dynamical
variable instead of the cosmic time $t$, we obtain the following
autonomous dynamical system,
\begin{align}\label{dynamicalsystemmultifluid}
& \frac{\mathrm{d}x_1}{\mathrm{d}N}=-9 A x_1^3 z+9 A x_1^2 z+3 x_1
\left(9 B x_2^3 z^2+x_2\right)-(c_1 x_2+c_2 x_1)\, ,
\\ \notag &
\frac{\mathrm{d}x_2}{\mathrm{d}N}=-3 x_2 \left(3 A x_1^2
z+1\right)+27 B x_2^4 z^2-27 B x_2^3
z^2+c_1 x_2+c_2 x_1+3 x_2^2\, , \\
\notag & \frac{\mathrm{d}z}{\mathrm{d}N}=\frac{1}{2} (-3) z
\left(-3 A x_1^2 z+9 B x_2^3 z^2+x_2\right)\, .
\end{align}
Finally, the total EoS parameter $w_{eff}$ which is defined as
$w_{eff}=\frac{p_{tot}}{\rho_t}$, in terms of the variables
(\ref{variablesofdynamicalsystem}) can be expressed as follows,
\begin{equation}\label{equationofstatetotal}
w_{eff}=-3 A x_1^2 z+9 B x_2^3 z^2-x_1\, ,
\end{equation}
where we took into account the Friedman constraint
(\ref{friedmannconstraint}). The dark energy EoS
(\ref{darkenergyeos}) for a single fluid Universe leads to
finite-time singularities, and the same applies for a system of
three cosmological fluids, namely consisting of baryons, dark
energy and dark matter fluid, as it was shown in Ref.
\cite{Odintsov:2018uaw}. However in Ref. \cite{Odintsov:2018uaw}
the dark matter fluid was assumed to be pressureless, hence it is
the purpose of this paper to investigate what happens in the case
that the dark matter component has the generalized EoS of the form
(\ref{darkmattereos}). In the following we shall use the
techniques firstly developed in \cite{goriely} in order to
investigate the singularity structure of the dynamical system, but
first let us demonstrate what is the structure of the phase space.

The standard techniques in order to reveal the phase space
structure of an autonomous dynamical system, is to rely on the
Hartman-Grobman theorem, which applies to hyperbolic fixed points.
Essentially, the Hartman-Grobman theorem states that given an
autonomous dynamical system of the form,
\begin{equation}\label{ds1}
\frac{\mathrm{d}\Phi}{\mathrm{d}t}=f(\Phi (t))\, ,
\end{equation}
where $f=(f_1,f_2,...,f_n)$, at the vicinity of a hyperbolic fixed
point, it is sufficient to study the linearized dynamical system,
\begin{equation}\label{loveisalie}
\frac{\mathrm{d}\Phi}{\mathrm{d}t}=\mathcal{J}(g)(\Phi)\Big{|}_{\Phi=\phi_*}
(\Phi-\phi_*)\, ,
\end{equation}
where $\mathcal{J}$ is the Jacobian,
\begin{equation}\label{jaconiab}
\mathcal{J}=\sum_i\sum_j\Big{[}\frac{\mathrm{\partial
f_i}}{\partial x_j}\Big{]}\, .
\end{equation}
Recall that a hyperbolic fixed point has a Jacobian matrix at a
hyperbolic fixed point, with eigenvalues that have non-zero real
part. In our case, the functions $f_i$ appearing in Eq.
(\ref{ds1}) are equal to,
\begin{align}\label{functionsfi}
& f_1=-9 A x_1^3 z+9 A x_1^2 z+3 x_1 \left(9 B x_2^3
z^2+x_2\right)-(c_1 x_2+c_2 x_1)\\
\notag & f_2=-3 x_2 \left(3 A x_1^2 z+1\right)+27 B x_2^4 z^2-27 B
x_2^3 z^2+c_1 x_2+c_2 x_1+3 x_2^2,\\
\notag & f_3=\frac{1}{2} (-3) z \left(-3 A x_1^2 z+9 B x_2^3
z^2+x_2\right)\, ,
\end{align}
and hence the Jacobian matrix is equal to,
\begin{align}\label{jacobianmatrix}
& \mathcal{J}=\\ \notag & \left(
\begin{array}{ccc}
 3 \left(9 B z^2 x_2^3+x_2+3 A x_1 (2-3 x_1) z\right)-c_2 & 81 B x_1 x_2^2 z^2-c_1+3 x_1 & 9 x_1 \left(6 B x_2^3 z-A (x_1-1) x_1\right) \\
 c_2-18 A x_1 x_2 z & 108 B z^2 x_2^3-81 B z^2 x_2^2+6 x_2+c_1-9 A x_1^2 z-3 & -9 \left(A x_1^2 x_2-6 B (x_2-1) x_2^3 z\right) \\
 9 A x_1 z^2 & -\frac{3}{2} z \left(27 B x_2^2 z^2+1\right) & -\frac{3}{2}  \left(27 B z^2 x_2^3+x_2-6 A x_1^2 z\right) \\
\end{array}
\right)\, .
\end{align}
By solving simultaneously the equations $(f_1,f_2,f_3)=(0,0,0)$,
we obtain the following fixed points for the dynamical system at
hand,
\begin{align}\label{fixedpointsc20}
& \phi_1=\{x_1\to 0,x_2\to 0,z\to 0\}
\\ \notag & \phi_2=\{x_1\to \frac{c_1 c_2+c_2
\sqrt{(-c_1+c_2+3)^2-12 c_2}-c_2^2+3 c_2}{6 c_2},x_2\to
\frac{1}{6} \left(-\sqrt{(-c_1+c_2+3)^2-12
c_2}-c_1+c_2+3\right),x_3\to 0\}, \\ \notag & \phi_3=\{x_1\to
\frac{c_1 c_2-c_2 \sqrt{(-c_1+c_2+3)^2-12 c_2}-c_2^2+3 c_2}{6
c_2},x_2\to \frac{1}{6} \left(\sqrt{(-c_1+c_2+3)^2-12
c_2}-c_1+c_2+3\right),z\to 0\}
 \, .
\end{align}
The fixed point $\phi_1$ does not seem to have any physical
significance at all, however the rest of the fixed points, namely
$\phi_2$ and $\phi_3$ do seem to have physical interest. Indeed,
the eigenvalues $j_1$, $j_2$ and $j_3$ of the Jacobian
(\ref{jacobianmatrix}) for the fixed point $\phi_1$ are,
\begin{equation}\label{eigenvaluesf1}
(j_1,j_2,j_3)=\Big{(}0,\frac{1}{2} \left(-\sqrt{(-c_1+c_2+3)^2-12
c_2}+c_1-c_2-3\right),\frac{1}{2} \left(\sqrt{(-c_1+c_2+3)^2-12
c_2}+c_1-c_2-3\right)\Big{)}\, ,
\end{equation}
which clearly shows that the fixed point is non-hyperbolic, so it
is impossible to apply the Hartman-Grobman theorem. Also let us
note that the EoS parameter (\ref{equationofstatetotal}) evaluated
at the fixed point $\phi_1$ reads $w_{eff}=0$, which clearly
describes a matter dominated epoch. However this case has no
physical interest so we do not further analyze this case, so let
us focus our analysis on the rest of the fixed points, $\phi_1$
and $\phi_2$. With regard to $\phi_2$, the eigenvalues $j_1$,
$j_2$ and $j_3$ of the Jacobian (\ref{jacobianmatrix}) are equal
to,
\begin{align}\label{eigenvaluesf2}
& j_1=\frac{1}{4} \left(\sqrt{(-c_1+c_2+3)^2-12
c_2}+c_1-c_2-3\right)  \\ \notag & j_2=\frac{-3 \sqrt{c_1^2-2 c_1
(c_2+3)+(c_2-3)^2}-c_1+c_2+3}{4}+\sqrt{\mathcal{S}(c_1,c_2)}\, ,
\\ \notag & j_3=\frac{-3
\sqrt{c_1^2-2 c_1
(c_2+3)+(c_2-3)^2}-c_1+c_2+3}{4}-\sqrt{\mathcal{S}(c_1,c_2)}\, ,
\end{align}
where the function $\mathcal{S}(c_1,c_2)$ stands for,
\begin{align}\label{sfunction}
& \mathcal{S}(c_1,c_2)=c_1^2+c_2 \sqrt{-2 (c_1+3)
c_2+(c_1-3)^2+c_2^2}+3 \sqrt{-2 (c_1+3)
c_2+(c_1-3)^2+c_2^2}+c_2^2+9 \\
\notag & -c_1\left(\sqrt{-2 (c_1+3) c_2+(c_1-3)^2+c_2^2}+2
c_2+6\right) \, .
\end{align}
Clearly, the fixed point $\phi_2$ is a hyperbolic one, for a wide
range of values of $c_1$ and $c_2$. A numerical analysis of the
parameter space $(c_1,c_2)$ shows that the fixed point $\phi_2$ is
unstable, due to the existence of positive eigenvalues of
$\mathcal{J}$. Let us focus on some specific cases with physical
interest, and the most interesting case is when $w_{eff}$ is close
to $-1$. The current observational bounds predict that
$w_{eff}=-1\pm 0.03$, so let us consider three cases, namely
$w_{eff}=-1$, $w_{eff}=-1.06$ and $w_{eff}=-1.03$. The case
$w_{eff}=-1$ is problematic, since we cannot have  $w_{eff}=-1$
for any value of $c_1$ and $c_2$, and therefore no de Sitter fixed
points of the dynamical system exist. With regard to the case
$w_{eff}=-1.03$, the parameters $c_1$ and $c_2$ must satisfy,
\begin{equation}\label{newfre}
c_2=-0.18+0.0566038c_1\, ,
\end{equation}
and as it can be shown, the eigenvalues of the Jacobian matrix
corresponding to the fixed point $\phi_2$, have for all the
numerical values of $c_1$ the structure $(a_{-},b_{+},c_{+})$,
with $a_->0$ and $b_{+},c_{+}<0$. Thus the fixed point $\phi_2$ is
hyperbolic and unstable, and it can also be seen the same applies
for the fixed point $\phi_3$, however we omit the details for
brevity. Finally the case $w_{eff}=-1.03$ is similar to the case
$w_{eff}=-1.06$, and in this case the parameters $c_1$ and $c_2$
must be related as follows,
\begin{equation}\label{newfre1}
c_2=-0.09+0.0291262c_1\, ,
\end{equation}
and the same procedure as above reveals that the fixed point
$\phi_2$ is an unstable hyperbolic fixed point.

\begin{figure}[h]
\centering
\includegraphics[width=20pc]{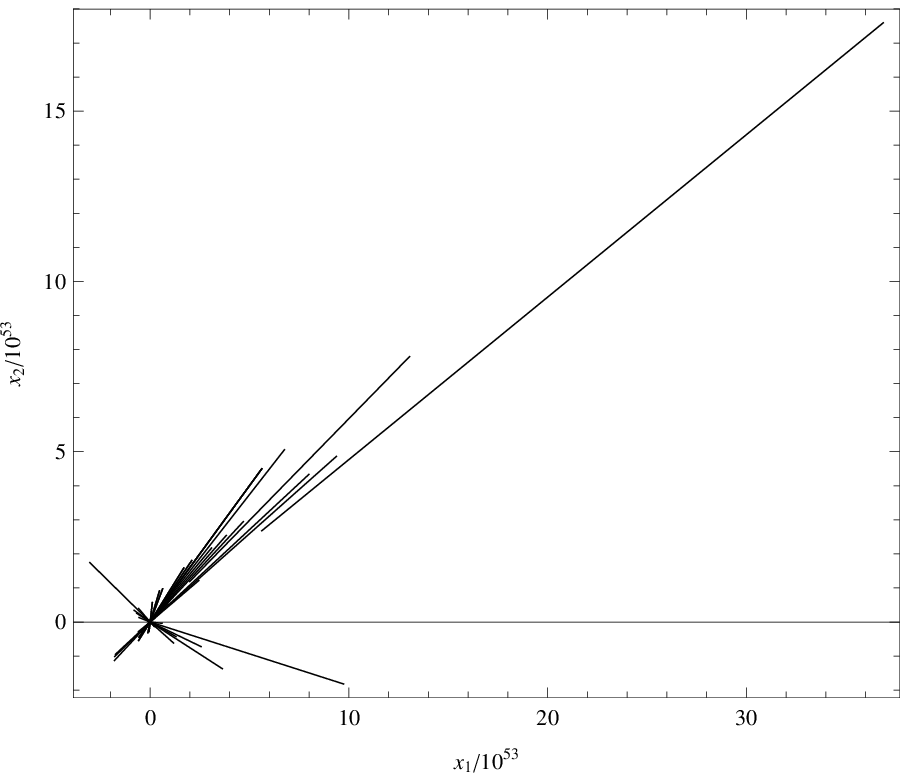}
\includegraphics[width=20pc]{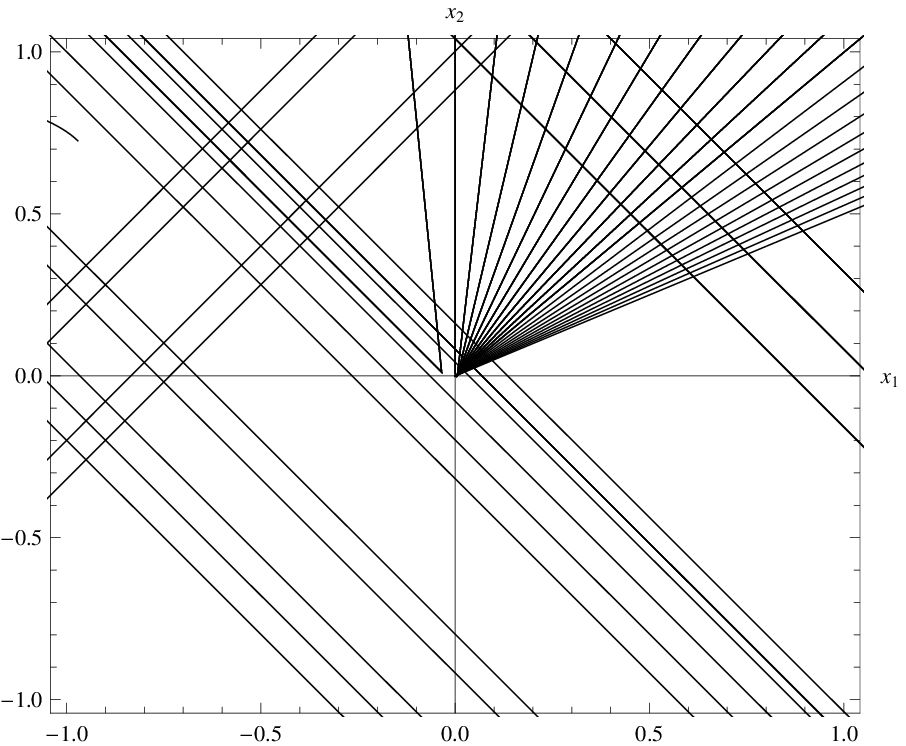}
\caption{{\it{The trajectories of the classical dark energy
superfluid dark matter interacting cosmological system for various
initial conditions, in the $x_1-x_2$ plane (left plot) and in the
$x_1-z$ plane (right plot).}}} \label{plot1}
\end{figure}

Our analysis so far clearly shows strong instabilities in the
classical interacting dark energy superfluid dark matter phase
space, and we also need to validate this numerically. To this end,
we solve numerically the dynamical system for various initial
conditions, and in Fig. \ref{plot1}, we present the phase space
trajectories in the $x_1-x_2$ plane (left plot) and in the $x_1-z$
plane. As it can be seen, there exist initial conditions which
make the corresponding trajectories blow up in the phase space.
This is a clear indication for the existence of finite-time
singularities in the cosmological system at hand. In Refs.
\cite{Odintsov:2018uaw,Odintsov:2018awm}, this issue was clearly
explained, and following the discussions in Refs.
\cite{Odintsov:2018uaw,Odintsov:2018awm}, a dynamical system
singularity, may be also a finite-time singularity for the
cosmological system. In fact, a singularity in the variables
$x_1$, $x_2$ and $z$ may correspond to a Type III singularity or a
Big Rip singularity of the cosmological system, we refer to Refs.
\cite{Odintsov:2018uaw,Odintsov:2018awm} for further details on
this, and to Ref. \cite{Nojiri:2005sx} for the classification of
finite-time singularities.

In principle, it is very hard to find finite-time singularities
using the dynamical systems approach, so one can rely on numerical
investigations. We need to note though the finite-time
singularities can be found by using some other analytic techniques
beyond the dynamical system approach, see for example
\cite{Barrow:2006xb,Ivanov:2011np}. In addition, in Refs.
\cite{Odintsov:2018uaw,Odintsov:2018awm} we used a powerful
technique, which we called ``Dominant Balance Analysis'', which
can be applied to any polynomial autonomous dynamical system.
Since the dynamical system (\ref{dynamicalsystemmultifluid}) is a
polynomial dynamical system, we shall also make use of the
dominant balance analysis in this paper too. The dominant balance
analysis is based on a theorem proved in Ref. \cite{goriely}, and
this analysis was also used in cosmological contexts in Ref.
\cite{barrowcotsakis}. It is worth recalling its basic features in
this paper too, which can be summarized in the following:

What we seek is proof that the variables of the dynamical system
(\ref{dynamicalsystemmultifluid}), namely, $x_1(N)$, $x_2(N)$ and
$z(N)$ become infinite at some finite-time instance $N=N_c$. So
consider the general dynamical system,
\begin{equation}\label{dynamicalsystemdombalanceintro}
\dot{x}=f(x)\, ,
\end{equation}
with $x$ being a vector of $R^n$, namely, $(x_1,x_2,...,x_n)$, and
$f(x)$ a vector of $R^n$ of the form
$f(x)=\left(f_1(x),f_2(x),...,f_n(x)\right )$, where the functions
$f_i(x)$ are strictly polynomials of the variables
$(x_1,x_2,...,x_n)$ of the dynamical system. At a finite-time
singularity a variable of the dynamical system will be have as
$(N-N_c)^{-p}$, with $p>0$, so the method of \cite{goriely} goes
as follows,
\begin{itemize}

\item Find all possible truncations of the function $f(x)$ in Eq.
(\ref{dynamicalsystemdombalanceintro}). A dominant truncation
controls the evolution of the dynamical system near finite-time
singularities, which we denote as $\hat{f}(x)$, hence the
dynamical system becomes,
\begin{equation}\label{dominantdynamicalsystem}
\dot{x}=\hat{f}(x)\, .
\end{equation}
Many different truncations may be found, but finding a consistent
truncation that satisfies the constraints we present now, is the
main aim of the method. For a truncation $\hat{f}(x)$, we write
the dynamical system variables in the following form,
\begin{equation}\label{decompositionofxi}
x_1(\tau)=a_1\tau^{p_1},\,\,\,x_2(\tau)=a_2\tau^{p_2},\,\,\,....,x_n(\tau)=a_n\tau^{p_n}\,
,
\end{equation}
so by assuming that the  solution $x$ can be written in
$\psi$-series form as functions of $\tau=N-N_c$, we substitute the
$x_i$'s from Eq. (\ref{decompositionofxi}) in Eq.
(\ref{dominantdynamicalsystem}). After that, one needs to equate
the exponents of the resulting polynomials, and therefore the
parameters $p_i$, $i=1,2,...,n$, are determined, which must be
real fractional numbers or integers. We form the vector
$\vec{p}=(p_1,p_2,...,p_n)$, and by equating the polynomials in
Eqs. (\ref{dominantdynamicalsystem}), (\ref{decompositionofxi}),
we obtain the coefficients $a_i$, from which we form the vector
$\vec{a}=(a_1,a_2,a_3,....,a_n)$. The set  $(\vec{a},\vec{p})\neq
0$ is called dominant balance.

\item  If $\vec{a}=(a_1,a_2,a_3,....,a_n)$ takes complex values
for some of the coefficients $a_i$'s, it is certain that the
dynamical system develops no finite-time singularities. On the
other hand if $\vec{a}=(a_1,a_2,a_3,....,a_n)$ takes real non-zero
values for all $a_i$'s, then the dynamical system develops
finite-time singularities.

\item  The singular and non-singular solutions found in the
previous step might correspond to general initial conditions or to
a limited set of initial conditions, and this must be ensured.
This feature can be validated by finding the Kovalevskaya matrix
$R$, which is defined as follows,
\begin{equation}\label{kovaleskaya}
R=\left(%
\begin{array}{ccccc}
  \frac{\partial \hat{f}_1}{\partial x_1} & \frac{\partial \hat{f}_1}{\partial x_2} & \frac{\partial \hat{f}_1}{\partial x_3} & ... & \frac{\partial \hat{f}_1}{\partial x_n} \\
  \frac{\partial \hat{f}_2}{\partial x_1} & \frac{\partial \hat{f}_2}{\partial x_2} & \frac{\partial \hat{f}_2}{\partial x_3} & ... & \frac{\partial \hat{f}_2}{\partial x_n} \\
  \frac{\partial \hat{f}_3}{\partial x_1} & \frac{\partial \hat{f}_3}{\partial x_2} & \frac{\partial \hat{f}_3}{\partial x_3} & ... & \frac{\partial \hat{f}_3}{\partial x_n} \\
  \vdots & \vdots & \vdots & \ddots & \vdots \\
  \frac{\partial \hat{f}_n}{\partial x_1} & \frac{\partial \hat{f}_n}{\partial x_2} & \frac{\partial \hat{f}_n}{\partial x_3} & ... & \frac{\partial \hat{f}_n}{\partial x_n} \\
\end{array}%
\right)-\left(%
\begin{array}{ccccc}
  p_1 & 0 & 0 & \cdots & 0 \\
  0 & p_2 & 0 & \cdots & 0 \\
  0 & 0 & p_3 & \cdots & 0 \\
  \vdots & \vdots & \vdots & \ddots & 0 \\
  0 & 0 & 0 & \cdots & p_n \\
\end{array}%
\right)\, ,
\end{equation}
which must be calculated at a non-zero balance $\vec{a}$ of the
previous step. Next one needs to calculate the eigenvalues of
$R(\vec{a})$, which must have the form $(-1,r_2,r_3,...,r_{n})$.

\item If $r_i>0$, $i=2,3,...,n$, then the solutions, singular or
not, are general solutions, which means that these correspond to a
general set of initial conditions. In the opposite case, the
solutions are not general and these correspond to a limited set of
initial conditions.

\end{itemize}
In the case at hand, we will apply the method we presented above,
having in mind that the variables $x_1$ and $x_2$ are required to
satisfy the Friedmann constraint (\ref{friedmannconstraint}), even
a singular point. The dynamical system
(\ref{dynamicalsystemmultifluid}) is written as
$\frac{\mathrm{d}\vec{x}}{\mathrm{d}N}=f(\vec{x})$, where
$\vec{x}$ is $\vec{x}=(x_1,x_2,z)$, and in addition the vector
function $f(x_1,x_2,z)$ is,
\begin{equation}\label{functionfmultifluidclassicalcase}
f(x_1,x_2,z)=\left(%
\begin{array}{c}
 f_1(x_1,x_2,z) \\
  f_2(x_1,x_2,z) \\
   f_3(x_1,x_2,z) \\
\end{array}%
\right)\, ,
\end{equation}
with the functions $f_i(x_1,x_2,z)$, $i=1,2,3$ being equal to,
\begin{align}\label{functionsficlassicalcase}
& f_1(x_1,x_2,z)= -9 A x_1^3 z+9 A x_1^2 z+3 x_1 \left(9 B x_2^3
z^2+x_2\right)-(c_1 x_2+c_2
x_1)\\
\notag & f_2(x_1,x_2,z)=-3 x_2 \left(3 A x_1^2 z+1\right)+27 B
x_2^4 z^2-27 B x_2^3 z^2+c_1 x_2+c_2 x_1+3 x_2^2, \\
\notag & f_3(x_1,x_2,z)=\frac{1}{2} (-3) z \left(-3 A x_1^2 z+9 B
x_2^3 z^2+x_2\right) \, .
\end{align}
A consistent truncation of
(\ref{functionfmultifluidclassicalcase}), is the following,
\begin{equation}\label{truncation1classicalcase}
\hat{f}(x_1,x_2,z)=\left(
\begin{array}{c}
 9 A x_1(N)^2 z(N) \\
 3 x_2(N)^2 \\
 \frac{1}{2} (-27) B x_2(N)^3 z(N)^3 \\
\end{array}
\right)\, ,
\end{equation}
and by following the steps of the dominant balance analysis
method, we easily obtain,
\begin{equation}\label{vecp1classicalcase}
\vec{p}=( -1, -1, 1 )\, ,
\end{equation}
and also,
\begin{align}\label{balancesdetails1classicalcase}
& \vec{a}=\Big{(}\frac{1}{9 \sqrt{2} A}, -\frac{1}{3},
-\sqrt{2}\Big{)} \, .
\end{align}
Clearly $\vec{a}$ cannot be complex for any values of $A$, hence
we definitely have singular solutions. It now remains to determine
whether the singular solutions are general or these correspond to
a limited set of initial conditions. To this end, we evaluate the
Kovalevskaya matrix $R$, which is,
\begin{equation}\label{kovalev1classicalcase}
R=\left(
\begin{array}{ccc}
 18 A x_1 z+1 & 0 & 9 A x_1^2 \\
 0 & 6 x_2+1 & 0 \\
 0 & \frac{1}{2} (-81) B x_2^2 z^3 & -\frac{1}{2} 81 B z^2 x_2^3-1 \\
\end{array}
\right)\, ,
\end{equation}
and when it is evaluated on $\vec{a}$, it is equal to,
\begin{equation}\label{ra}
R(\vec{a})=\left(
\begin{array}{ccc}
 -1 & 0 & \frac{1}{18 A} \\
 0 & -1 & 0 \\
 0 & 9 \sqrt{2} B & 3 B-1 \\
\end{array}
\right)\, ,
\end{equation}
the eigenvalues of which are,
\begin{equation}\label{eigenvalues1classicalcase}
(r_1,r_2,r_3,r_4)=(-1,-1,-1 + 3 B)\, .
\end{equation}
Due to the fact that it contains negative eigenvalues, this
indicates that the singular solutions found for the classical dark
energy superfluid dark matter system are not general, which means
that these correspond to a limited set of initial conditions.

In conclusion, our analysis of the phase space of the classical
dark energy superfluid dark matter cosmological system indicates
that no de Sitter fixed points exist, and the rest of the fixed
points, which can be of various physical forms, for example
radiation or matter domination ones, are strongly unstable. Also
the dominant balance analysis we performed showed that there exist
singular solutions which are not general solutions though, which
means that these correspond to a limited set of initial
conditions. What now remains is to investigate whether the LQC
effects affect the phase space structure of the dark energy-dark
matter system. This is the subject of the next section.

\section{The Loop Quantum Cosmology Framework and Interacting Dark Energy with Superfluid Dark Matter}

Having discussed the classical phase space of the coupled dark
energy superfluid dark matter system, in this section we shall
thoroughly investigate the LQC phase space of this cosmological
system. We shall focus on the existence of stable attractors, with
particular emphasis given on de Sitter attractors, which are
relevant for the late-time acceleration. Also we shall investigate
whether the singular solutions of the classical dynamical system
persist in the LQC case too. However it is worth discussing
briefly what superfluid dark matter brings along and why it is
worth studying. Superfluid dark matter was introduced in Ref.
\cite{Berezhiani:2015bqa}, and further developed in Refs.
\cite{Berezhiani:2015pia,Hodson:2016rck,Berezhiani:2017tth}. The
dark matter superfluid model of Ref. \cite{Berezhiani:2015bqa} has
the appealing property of reproducing the $\Lambda$-Cold Dark
Matter model at large scales, and it simultaneously matches the
phenomenological predictions of Modified Newtonian Dynamics on
Galactic scales. In the theory of Ref. \cite{Berezhiani:2015bqa}
consists of axion-like particles with eV-scale mass, which
strongly interact between them. The effects of the superfluid dark
matter condensates are apparent only on galactic scales, but not
when galactic clusters are considered. The resulting theory has
many appealing observational features, like for example it
explains the low-density  vortices in galaxies, it also explains
the infall dependent and phonon speed of sound low-density
vortices in galaxies. In addition, bullet-like clusters are also
very well fitted by the theory which leads to distinct mass peaks,
and furthermore the the superfluid effective theory has
similarities with a unitary Fermi Gas. For further discussions and
phenomenological implications on superfluid dark matter, we refer
the reader to Refs.
\cite{Berezhiani:2015pia,Hodson:2016rck,Berezhiani:2017tth}.

Before we start, we shall briefly review some essential
information of LQC, and for details, the reader is referred to
Refs.
\cite{LQC1,LQC3,LQC4,LQC5,Salo:2016dsr,Xiong:2007cn,Amoros:2014tha,Cai:2014zga,deHaro:2014kxa,Kleidis:2018plu,Kleidis:2017ftt}.
We focus on holonomy corrected LQC, in the context of which, the
spacetime is discrete, and the Hamiltonian of the theory is
written in terms of the holonomies
$h_j=e^{-\frac{i\lambda\sigma_j}{2}}$, where $\sigma_j$ denote the
Pauli matrices, and it is equal to,
\begin{equation}\label{hamiltonianlqc}
\mathrm{H}_{LQC}=-\frac{2V}{\gamma^3\lambda^3}\Sigma_{i,j,k}\epsilon^{ijk}\mathrm{Tr}[h_i(\lambda)h_j(\lambda)h_i^{-1}(\lambda)\{h_k^{-1},V\}]+\rho
V\, .
\end{equation}
In the Hamiltonian (\ref{hamiltonianlqc}), the parameter $\gamma$
is $\gamma=0.2375$ and it is the Barbero-Immirzi parameter, and
also the parameter $\lambda$ is
$\lambda=\sqrt{\frac{\sqrt{3}}{2}\gamma}=0.3203$, and it has
dimension of length. Furthermore, $V$ stands for the spacetime
volume, which for the flat FRW metric reads $V=a^3$, and finally
$\rho$ denotes the Universe's total energy density. Moreover, the
dynamical variable $\beta$ is the canonical conjugate variable
$V$, and these two have the following Poisson bracket
$\{\beta,V\}=\frac{\gamma}{2}$. The trace of the Hamiltonian can
easily be evaluated and it reads,
\begin{equation}\label{hamiltonianlqc2}
\mathrm{H}_{LQC}=-3V\frac{\sin^2(\lambda \beta)}{\gamma^2\lambda^2}+\rho V\, ,
\end{equation}
and by also taking into account the Hamiltonian constraint
$\mathrm{H}_{LQC}=0$,  we have,
\begin{equation}\label{hcqrefre1}
\frac{\sin^2(\lambda \beta)}{\gamma^2\lambda^2}=\frac{\rho}{3}\, ,
\end{equation}
which is the LQC version of the Friedmann equation. By taking into
account the equation
$\dot{V}=\{V,\mathrm{H}_{LQC}\}=-\frac{\gamma}{2}\frac{\partial
\mathrm{H}_{LQC}}{\partial \beta}$, then by combining the above we
obtain,
\begin{equation}\label{frwhceqn1}
H=\frac{\sin(\lambda \beta)}{\gamma \lambda}\, ,
\end{equation}
which can be written as,
\begin{equation}\label{frwhceqn2}
\beta=\frac{\arcsin (2\lambda \gamma H)}{2\lambda}\, .
\end{equation}
By using Eqs. (\ref{frwhceqn2}) and (\ref{hcqrefre1}), we obtain,
\begin{equation}\label{frwhceqn3}
\frac{\sin^2(\lambda \frac{\arcsin (2\lambda \gamma
H)}{2\lambda})}{\gamma^2\lambda^2}=\frac{\rho}{3}\, ,
\end{equation}
and after some algebraic manipulations, we obtain the final form
of the LQC Friedmann equation, which has the following form,
\begin{equation}\label{lqcfriedmannequation}
H^2=\frac{\rho}{3}\left( 1-\frac{\rho}{\rho_c}\right)\, .
\end{equation}
The critical density parameter $\rho_c$ is of great importance,
due to the fact that it is the maximum energy density of the LQC
Universe, and it is equal to
$\rho_c=\frac{3}{\gamma^2\lambda^2}\cong 258$. Notably, the limit
$\rho_c\to \infty$ in Eq. (\ref{lqcfriedmannequation}) restores
the classical Friedmann equation (\ref{flateinstein}).

\begin{figure}[h]
\centering
\includegraphics[width=20pc]{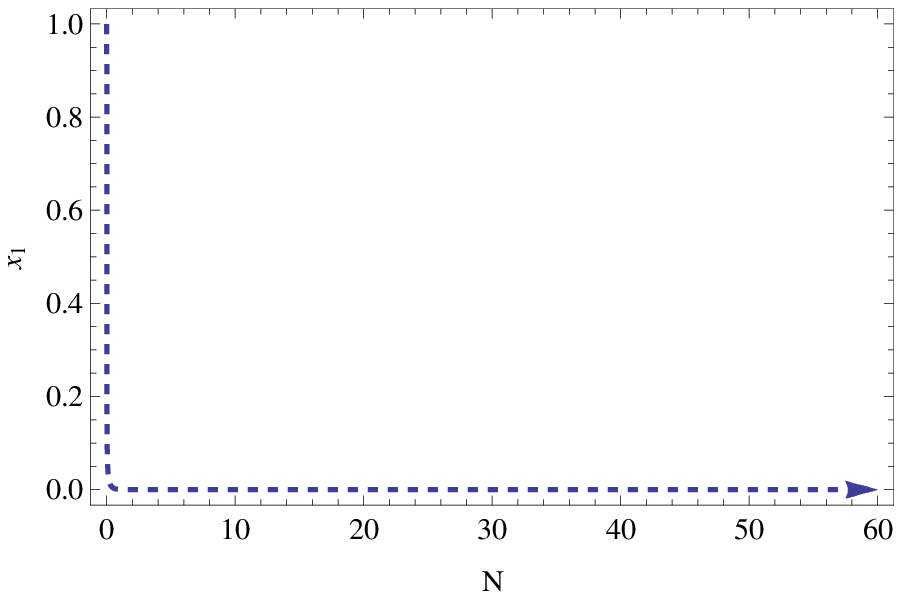}
\includegraphics[width=20pc]{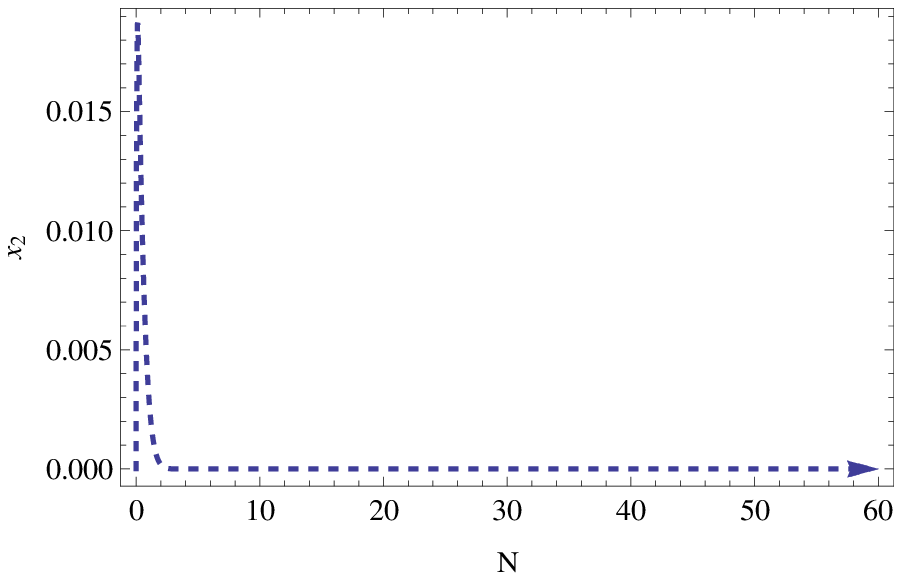}
\includegraphics[width=20pc]{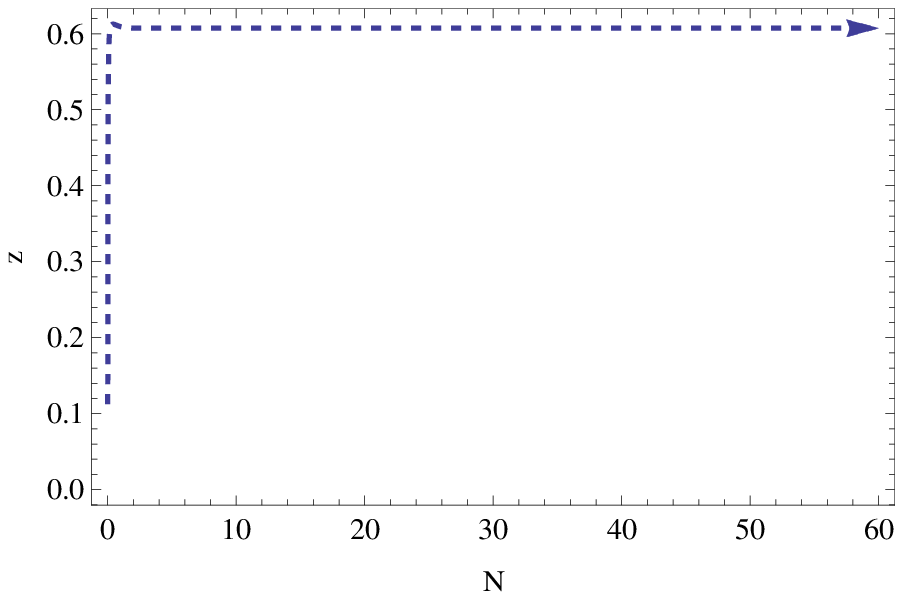}
\caption{{\it{The behavior of $x_1(N)$, $x_2(N)$ and $z(N)$ as
functions of $N$, for $w_d=0$ and $A<0$, $B>0$, $c_1,c_2>0$. As it
can be seen, a stable fixed point is reached quite fast.}}}
\label{plotasx}
\end{figure}

Let us now proceed in finding the LQC dynamical system of the
coupled dark energy superfluid dark matter fluid for the flat FRW
metric (\ref{frw}), so let us write the LQC Friedman equation in
the following form,
\begin{equation}\label{flateinsteinlqccase}
H^2=\frac{\kappa^2\rho_{tot}}{3}\left(
1-\frac{\rho_{tot}}{\rho_c}\right)\, ,
\end{equation}
with $\rho_{tot}$ being in this case too the total energy density
$\rho_{tot}=\rho_d+\rho_m$. The continuity equations for the dark
energy and the dark matter fluids remain the same as in Eq.
(\ref{continutiyequations}) and also the interaction term $Q$ is
assumed to have the form (\ref{qtermform}). Upon differentiation
of Eq. (\ref{flateinsteinlqccase}) with respect to the cosmic
time, in conjunction with Eq. (\ref{continutiyequations}), we get,
\begin{equation}\label{derivativeofhlqccase}
\dot{H}=-\frac{\kappa^2}{2}\left(\rho_m+\rho_d+p_{tot}\right)\left(
1-2\frac{\rho_m+\rho_d}{\rho_c}\right)\, ,
\end{equation}
where $p_{tot}$ is the total pressure which is equal to
$p_{tot}=p_d+p_m$. In addition, for the purposes of this section,
we assume that the dark energy EoS of equation
(\ref{darkenergyeos}), has the following form,
\begin{equation}\label{darkenergyeoslqccaselqccase}
p_d=-\rho_d-w_d\rho_d-\frac{A}{\rho_c}\rho_d^2\, ,
\end{equation}
where $w_d$ is a free parameter of the theory, and also that the
dark matter EoS has the superfluid form of the form,
\begin{equation}\label{darkmattereoslqccase}
p_m=\frac{B}{\rho_c^2}\rho_m^3\, ,
\end{equation}
where $A$ and $B$ are dimensionless constant parameters. Our aim
is to construct a polynomial autonomous dynamical system for the
cosmological system at hand, analogous to the one appearing in Eq.
(\ref{dynamicalsystemmultifluid}), so we shall choose the
dimensionless variables of the dynamical system to have the
following form in this case,
\begin{equation}\label{variablesofdynamicalsystemlqccase}
x_1=\frac{\kappa^2\rho_d}{3H^2},\,\,\,x_2=\frac{\kappa^2\rho_m}{3H^2},\,\,\,z=\frac{H^2}{\kappa^2\rho_c}\,
.
\end{equation}
In view of Eq. (\ref{flateinsteinlqccase}), the variables $x_i$,
$i=1,2$ and $z$ satisfy the LQC version of the Friedmann
constraint, which is,
\begin{equation}\label{friedmannconstraintlqccase}
x_1+x_2-z\left(x_1+x_2+x_3\right)^2=1\, .
\end{equation}
Moreover, the total EoS parameter $w_{eff}$ has in the LQC case
the following form,
\begin{equation}\label{equationofstatetotallqccase}
w_{eff}=\frac{-3 A x_1^2 z+9 B x_2^3 z^2-w_d x_1-x_1}{x_1+x_2}\, .
\end{equation}
\begin{figure}[h]
\centering
\includegraphics[width=20pc]{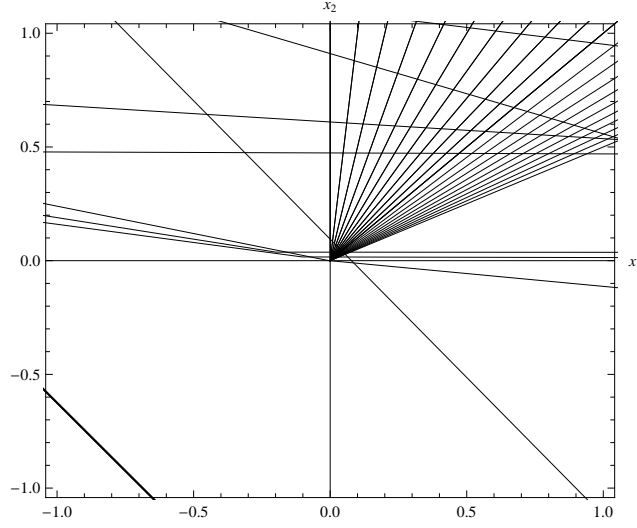}
\caption{{\it{The trajectories of the LQC dark energy superfluid
dark matter interacting cosmological system for various initial
conditions, in the $x_1-x_2$ plane.}}} \label{plot2}
\end{figure}
By combining Eqs. (\ref{flateinsteinlqccase}),
(\ref{derivativeofhlqccase}), (\ref{continutiyequations}), and
(\ref{variablesofdynamicalsystemlqccase}), and also by using the
$e$-foldings number as a dynamical variable, after some extensive
algebraic manipulations, we obtain the dynamical system,
\begin{align}\label{dynamicalsystemmultifluidlqccase}
& \frac{\mathrm{d}x_1}{\mathrm{d}N}=54 A x_1^4 z^2+54 A x_1^3 x_2
z^2-9 A x_1^3 z+9 A x_1^2 z-162 B x_1^2 x_2^3
z^3-162 B x_1 x_2^4 z^3+27 B x_1 x_2^3 z^2\\
\notag &-3 (c_1 x_2+c_2 x_1)+18 w_d x_1^3 z+18 w_d x_1^2 x_2 z-3
w_d
x_1^2+3 w_d x_1-18 x_1^2 x_2 z\\
\notag &-18 x_1 x_2^2 z+3 x_1 x_2 \, ,
\\ \notag &
\frac{\mathrm{d}x_2}{\mathrm{d}N}=54 A x_1^3 x_2 z^2+54 A x_1^2
x_2^2 z^2-9 A x_1^2 x_2 z-162 B x_1 x_2^4 z^3-162 B x_2^5 z^3+27 B
x_2^4 z^2-27 B x_2^3 z^2\\ \notag &+3 (c_1 x_2+c_2 x_1)+18
w_d x_1^2 x_2 z+18 w_d x_1 x_2^2 z-3 w_d x_1 x_2-18 x_1 x_2^2 z\\
\notag &-18 x_2^3 z+3 x_2^2-3 x_2
 \, ,
\\ \notag & \frac{\mathrm{d}z}{\mathrm{d}N}=-54 A x_1^3 z^3-54 A x_1^2 x_2 z^3+9 A x_1^2 z^2
+162 B x_1 x_2^3 z^4\\ \notag &+162 B x_2^4 z^4-27 B x_2^3 z^3 -18
w_d x_1^2 z^2-18 w_d x_1
x_2 z^2+3 w_d x_1 z+18 x_1 x_2 z^2\\
\notag &+18 x_2^2 z^2-3 x_2 z \, ,
\end{align}
where we also took into account Eq. (\ref{additionalterms}). The
dynamical system (\ref{dynamicalsystemmultifluidlqccase})
describes the LQC coupled dark energy superfluid dark matter
cosmological system, and it is an autonomous polynomial dynamical
system, which we now study thoroughly. As we will demonstrate, the
phase space structure of the LQC cosmological system is very rich.
We start off with the fixed points, which are,
\begin{align}\label{fixedpointsc20lqccase}
& \phi_1^*=\{x_1\to 0,x_2\to 0,z\to z\}, \\ \notag &
\phi_2^*=\{x_1\to \frac{\mathcal{S}+c_1-c_2+w_d+1}{2
(w_d+1)},x_2\to \frac{-\mathcal{S}-c_1+c_2+w_d+1}{2 (w_d+1)},z\to 0\},\\
\notag & \phi_3^*=\{x_1\to \frac{-\mathcal{S}+c_1-c_2+w_d+1}{2
(w_d+1)},x_2\to \frac{\mathcal{S}-c_1+c_2+w_d+1}{2 (w_d+1)},z\to
0\}\, ,
\end{align}
where $\mathcal{S}$ stands for,
\begin{equation}\label{newasx1oen}
\mathcal{S}=\sqrt{(-c_1+c_2+w_d+1)^2-4 c_2 (w_d+1)}\, .
\end{equation}
Apart from the above, it can be shown that there exist other
classes of fixed points, which are determined by the simultaneous
validity of various algebraic equations, which are too lengthy to
quote here. We denote these fixed points $\phi_*^{\dag}$. Actually
the class of fixed points $\phi_*^{\dag}$ are more interesting
phenomenologically, since some of these are stable when the de
Sitter cosmology is considered. In principle, by requiring
$w_{eff}=0$ or $w_{eff}=1/3$, we can obtain the radiation and
matter domination fixed points, however we shall not be interested
in these fixed points, but we emphasize on de Sitter fixed points,
which as we show these exist. Let us start with the fixed point
$\phi_1^*$ which is unphysical since the total EoS parameter is
infinite, so let us proceed in the study of the fixed point
$\phi_2^*$. In this case we require that $w_{eff}=-1$, which
describes a de Sitter evolution. The condition $w_{eff}=-1$ is
satisfied when,
\begin{equation}\label{condtiondesitter}
c_2=w_d-c_1 w_d\, ,
\end{equation}
and this covers the fixed point $\phi_3^*$ too. So let us
investigate the stability of these de Sitter vacua, and by
calculating the Jacobian matrix $\mathcal{J}$ and the
corresponding eigenvalues, the resulting picture is that these
vacua are hyperbolic fixed points, however unstable for all the
values of $w_d$ and $c_1$. Note that due to the extended and
complicated form of the Jacobian matrix and of the resulting
eigenvalues, we do not quote these here. Let us now investigate
the stability of the class of fixed points $\phi_*^{\dag}$, so we
perform a numerical analysis for this case. The resulting picture
is that for $w_d=0$ and for $A<0$, $B>0$, $c_1,c_2>0$ and for
small negative values of $x_2$, some de Sitter vacua of
$\phi_*^{\dag}$ are hyperbolic fixed points with negative real
parts of the eigenvalues of the Jacobian matrix, hence stability
is ensured. In order to have a more clear picture of how the
trajectories behave in the phase space, we shall perform a
numerical investigation. As we shall see, the trajectories in the
phase space approach a stable fixed point, and this can be seen in
the numerical analysis of the phase space trajectories which we
present in Fig. \ref{plot2}. As it can be seen in Fig. \ref{plot2}
where we plot the trajectories in the $x_1-x_2$ plane, and as it
can be seen a stable equilibrium is reached by the trajectories
for various initial conditions. Also, in Fig. \ref{plotasx} we
plot the behavior of $x_1(N)$, $x_2(N)$ and $z(N)$ as functions of
$N$, for the values of the free parameters chosen as $w_d=0$ and
$A<0$, $B>0$, $c_1,c_2>0$, and as it can be seen a stable fixed
point is reached.
An important feature in order for stable and physically acceptable
fixed points to occur, is that $A$ must be negative and also
$c_1,c_2>0$. If $c_1,c_2<0$ or if some of these are negative, the
final value of the variable $x_2$ is negative, so this is not
physically acceptable.

\begin{figure}[h]
\centering
\includegraphics[width=20pc]{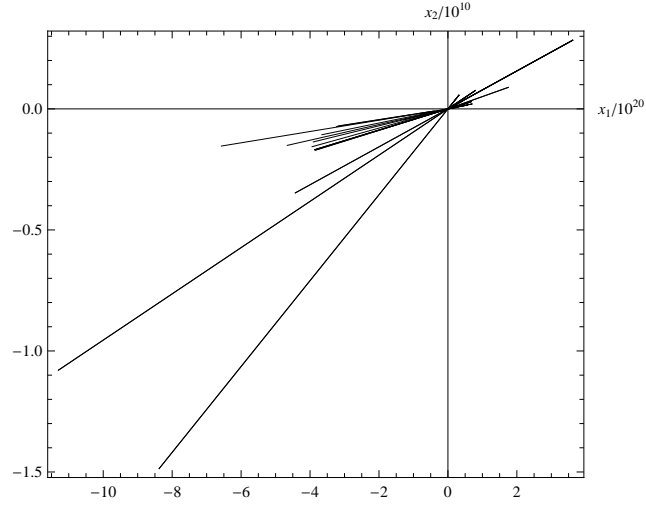}
\caption{{\it{The singular trajectories of the LQC dark energy
superfluid dark matter interacting cosmological system for various
initial conditions, in the $x_1-x_2$ plane.}}} \label{plot3}
\end{figure}

A deeper analysis revealed strong singular solutions, for which
the variables blow-up, see for example Fig. \ref{plot3}, where we
plot the trajectories in the $x_1-x_2$ plane.  This mean that
there exist initial conditions which make the variables blow-up
strongly, and thus we shall use the dominant balance analysis we
presented briefly in the previous section, in order to investigate
whether the singular solutions correspond to general initial
conditions or to a limited set of initial conditions.
To this end, we can write the dynamical system as
(\ref{dynamicalsystemmultifluidlqccase})
$\frac{\mathrm{d}\vec{x}}{\mathrm{d}N}=f(\vec{x})$, with the
vector $\vec{x}$ being of the form $\vec{x}=(x_1,x_2,x_3,z)$, and
in addition the vector function $f(x_1,x_2,z)$ being defined in
the following way,
\begin{equation}\label{functionfmultifluidlqccase}
f(x_1,x_2,z)=\left(%
\begin{array}{c}
 f_1(x_1,x_2,z) \\
  f_2(x_1,x_2,z) \\
   f_3(x_1,x_2,z) \\
\end{array}%
\right)\, ,
\end{equation}
with the functions $f_i(x_1,x_2,z)$, $i=1,2$ being equal to,
\begin{align}\label{functionsfilqccase}
& f_1(x_1,x_2,z)=54 A x_1^4 z^2+54 A x_1^3 x_2 z^2-9 A x_1^3 z+9 A
x_1^2 z-162 B x_1^2 x_2^3
z^3-162 B x_1 x_2^4 z^3+27 B x_1 x_2^3 z^2\\
\notag &-3 (c_1 x_2+c_2 x_1)+18 w_d x_1^3 z+18 w_d x_1^2 x_2 z-3
w_d
x_1^2+3 w_d x_1-18 x_1^2 x_2 z\\
\notag &-18 x_1 x_2^2 z+3 x_1 x_2 \, ,
\\ \notag &
f_2(x_1,x_2,z)=54 A x_1^3 x_2 z^2+54 A x_1^2 x_2^2 z^2-9 A x_1^2
x_2 z-162 B x_1 x_2^4 z^3-162 B x_2^5 z^3+27 B x_2^4 z^2-27 B
x_2^3 z^2\\ \notag &+3 (c_1 x_2+c_2 x_1)+18
w_d x_1^2 x_2 z+18 w_d x_1 x_2^2 z-3 w_d x_1 x_2-18 x_1 x_2^2 z\\
\notag &-18 x_2^3 z+3 x_2^2-3 x_2\, ,  \\ \notag &
f_3(x_1,x_2,z)=-54 A x_1^3 z^3-54 A x_1^2 x_2 z^3+9 A x_1^2 z^2
+162 B x_1 x_2^3 z^4\\ \notag &+162 B x_2^4 z^4-27 B x_2^3 z^3 -18
w_d x_1^2 z^2-18 w_d x_1
x_2 z^2+3 w_d x_1 z+18 x_1 x_2 z^2\\
\notag &+18 x_2^2 z^2-3 x_2 z \, .
\end{align}
A consistent truncation of the function $f(x_1,x_2,z)$ appearing
in Eq. (\ref{functionfmultifluidlqccase}), is the following,
\begin{equation}\label{truncation1lqccase}
\hat{f}(x_1,x_2,z)=\left(
\begin{array}{c}
 3 x_1(N) x_2(N) \\
-3 w_d x_1(N) x_2(N) \\
162 B x_1 x_2^3 z^4 \\
\end{array}
\right)\, .
\end{equation}
The vector $\vec{p}$ can easily be found by applying the method of
the previous section, so we obtain,
\begin{equation}\label{vecp1lqccase}
\vec{p}=( -1, -1, 1 )\, ,
\end{equation}
and accordingly, the vector $\vec{a}$ reads,
\begin{align}\label{balancesdetails1lqccase}
& \vec{a}=\Big{(}\frac{1}{3 w_d}, -\frac{1}{3},
-\frac{\sqrt{3}{w_d}}{\sqrt{3}{2} \sqrt{3}{B}} \Big{)}\, .
\end{align}
The vector $\vec{a}$ can be complex for $B<0$ and real for $B>0$,
so we shall investigate both cases of the sign of the parameter
$B$. The Kovalevskaya matrix $R$ for the truncation
(\ref{truncation1lqccase}) reads,
\begin{equation}\label{kobvalev1lqccasea1}
R(\vec{a})=\left(
\begin{array}{ccc}
 3 x_2+1 & 3 x_1 & 0 \\
 -3 w_d x_2 & 1-3 w_d x_1 & 0 \\
 162 B x_2^3 z^4 & 486 B x_1 x_2^2 z^4 & 648 B x_1 x_2^3 z^3-1 \\
\end{array}
\right)\, ,
\end{equation}
and therefore, when it is evaluated on $\vec{a}$ it reads,
\begin{equation}\label{kobvalev1lqccase}
R(\vec{a})=\left(
\begin{array}{ccc}
 0 & \frac{1}{w_d} & 0 \\
 w_d & 0 & 0 \\
 -\frac{3 w_d^{4/3}}{\sqrt[3]{2} \sqrt[3]{B}} & \frac{9 \sqrt[3]{w_d}}{\sqrt[3]{2} \sqrt[3]{B}} & 3 \\
\end{array}
\right)\, .
\end{equation}
The eigenvalues of the above matrix are,
\begin{equation}\label{eigenvalues1lqccase}
(r_1,r_2,r_3,r_4)=(-1,1,3)\, ,
\end{equation}
regardless of the choice of the parameter $B$. Therefore, the
resulting picture is quite interesting, since it validates our
earlier numerical analysis considerations which indicated that for
positive values of $B$, singular solutions exist. Indeed, the
eigenvalues (\ref{eigenvalues1lqccase}) indicate that for positive
$B$, $\vec{a}$ is real and therefore there exist singular
solutions which correspond to general initial conditions (due to
the form of the eigenvalues (\ref{eigenvalues1lqccase})), a
feature that we also demonstrated in Fig. \ref{plot3}. Also, for
negative $B$, our analysis shows that no general singular
solutions exist, but in this case, no stable de Sitter equilibria
exist in the phase space of the LQC system. With regard to the
relation of the dynamical system singular solutions (for $B<0$)
with the finite time singularities, when $x_1$ and $x_2$ actually
blow up, the finite-time singularities may be of Big Rip or even
Type II or Type III type depending on the value of the parameter
$z$. Due to the form of the vector $\vec{p}$, we may conclude that
in the case $B<0$, both $x_1$ and $x_2$ blow up in the phase
space, and therefore $z$ is finite, which means that the
singularities are of Type III, according to the classification of
Ref. \cite{Nojiri:2005sx}. Finally, we need to question the
physical significance of the cases with $B<0$, since in this case,
the dark matter EoS (\ref{darkmattereos}) would describe a
negative pressure fluid, which is highly unlikely however. Hence
the case $B<0$ is rather physically unappealing.

In conclusion, the LQC extended coupled dark energy superfluid
dark matter cosmological system has interesting phase features,
which we list in Table \ref{table1}, along with the classical
cosmological system and we discuss here in brief. The LQC system
has many de Sitter vacua, which for $B>0$ can be stable de Sitter
equilibria. Also for $B>0$, general singular solutions exist in
the phase space, along with stable de Sitter equilibria, which
means that apart from the set of initial conditions which may lead
to the stable de Sitter fixed points, there exist general initial
conditions which lead to singular solutions. For $B<0$ however, we
demonstrated that no singular solutions exist in the phase space.
This behavior is to be contrasted with the classical case, where
no de Sitter fixed points existed and also only a limited set of
initial conditions leaded to singular solutions.
\begin{table*}[h]
\small \caption{\label{table1}Phase Space Structure of the
classical and LQC coupled dark energy superfluid dark matter
system.}
\begin{tabular}{@{}crrrrrrrrrrr@{}}
\tableline \tableline \tableline
 Theoretical Framework &$\,\,\,\,$ Existence and Stability of de Sitter Fixed points & Singular Solutions
\\\tableline
Classical Case: & No de Sitter fixed Points$\,\,\,$$\,\,\,$ $\,\,\,$  & Non general singular solutions \\
\\\tableline
LQC case: & Stable de Sitter fixed points for $B>0$$\,\,\,$$\,\,\,$$\,\,\,$ &  General singular solutions for $B>0$ \\
LQC case: & Non-stable de Sitter fixed points for $B<0$$\,\,\,$$\,\,\,$$\,\,\,$ &  No singular solutions for $B<0$\\
\\\tableline
 \tableline
\end{tabular}
\end{table*}

Before closing, an important question arises, related to the era
were LQC effects are expected to be found. Particularly it is
known that the LQC effects should be found in early-time eras,
were gravity is expected to be considerably strong. In the case at
hand, when interacting dark energy-dark matter fluids are
considered in the context of LQC, we found stable de Sitter
attractors for the theory. These attractor solutions should be
interpreted correctly, mathematically these solutions exist, but
the question is what do they represent? Obviously, this question
cannot be answered by using only the fixed points of the dynamical
system, since there is no obvious answer on when these fixed
points are reached from the dynamical system. Our numerical
analysis showed that these fixed points are reached quite fast,
before the first sixty e-foldings, see for example Fig.
\ref{plotasx}. Thus these could be viewed as possibly some
early-time inflationary attractors, which are reached quite fast
from the dynamical system. Also the existence of singular
solutions in this case, shows that there exist initial conditions
in the system that will generate finite-time singularities, which
possibly can be related to the late-time era, since the early-time
era is quite stable and de Sitter like. To our opinion, in order
to be accurate, the quantum equations of motion will be those of
LQC only when the energy density of the Universe is quite close to
$\rho_c$, otherwise one effectively has the classical equations of
motion. Therefore, the possible interpretation of the results is
that the dark energy fluid coupled with the dark matter fluid
gives some early inflationary attractors in the theory, when LQC
are strong, but as the Universe evolves, and the cosmological
equations become effectively classical, the coupled superfluid
dark matter-dark energy system has no de Sitter attractor
solutions. Thus this makes superfluid dark matter quite important
for early-time considerations, at least when dark energy is
represented by a fluid coupled to the dark matter fluid. We need
to note that it is quite interesting to check whether the
superfluid composed by axion-like particles has strong effects if
it is considered in the context of other theories, like for
example modified gravity. Work is in progress in this research
line.

\section{Conclusions}

In this paper we investigated the classical and the LQC phase
space of a cosmological system consisting of two interacting dark
fluids, namely the dark energy fluid and the superfluid dark
matter fluid. For the dark matter fluid we assumed that the EoS
describes superfluid dark matter, so the fluid has non-zero
pressure, in contrast to cold dark matter which is pressureless.
With regard to the classical case, we investigated the existence
and stability of cosmological fixed points, and as we demonstrated
there exist matter and radiation domination fixed points which are
unstable, however no de Sitter fixed points occur in the phase
space for any value of the free parameters. Due to the structure
of the dynamical system, which was a polynomial autonomous
dynamical system, we were able to perform a dominant balance
analysis, which revealed that no general singular solutions exist,
however singular solutions corresponding to a limited set of
initial conditions exist. With regard to the LQC phase space, the
situation is much more interesting due to the fact that the phase
space contains much more physical structures in comparison to the
classical one. Specifically, in this case stable de Sitter fixed
points occur when the free parameter $B$, appearing in the
superfluid dark matter EoS, is positive. In this case too, the
phase space has also singular solutions, which correspond to a
general set of initial conditions, so these are general solutions.
In addition, the singular solutions of the dynamical system
correspond to Type III physical finite-time singularities. We
believe that we covered the most general case for the coupled
system of dark energy and superfluid dark matter system, and the
only modification which could be done in principle is to use an
EoS for the dark energy fluid of the form $p_d=-\rho_d-A\rho_d^n$,
$n>2$, which could have strong effects in the singularity
occurrence phenomenon. We defer this task to a future work.
Different EoS for dark energy, such as logarithmic
\cite{Odintsov:2018obx} or even Chaplygin gas, like in the form
used in \cite{Bamba:2012cp,Bento:2002ps,Bilic:2001cg} (see also
\cite{Khurshudyan:2018kfk} where a non-trivial interacting varying
Chaplygin gas EoS is used with tachyonic matter, which would
correspond to $B<0$ in the case studied in this paper), would make
the form of the dynamical system quite complicated, since it would
cease to be a polynomial dynamical system, so the method of
dominant balance analysis which we used in this paper, would be
inapplicable in this case. Such a task would require different
approaches, and work is in progress along this research line.

\end{document}